\documentclass[
    aps,
    prb,
    twocolumn,
    superscriptaddress,
    showpacs,
    floatfix
]{revtex4-2}

\usepackage{amsmath}
\usepackage{amssymb}
\usepackage{cancel}
\usepackage{bm}
\usepackage{xcolor}
\usepackage{dcolumn}
\usepackage{eucal}
\usepackage{mathrsfs}
\usepackage{graphicx}
\usepackage{booktabs}
\usepackage{adjustbox}
\usepackage{braket}
\usepackage{xfrac}
\usepackage{textcomp}
\usepackage{times}
\usepackage[toc,page]{appendix}
 \usepackage{ragged2e} 
\usepackage[
    breaklinks,
    colorlinks=true,
    linkcolor=red,
    urlcolor=cyan,
    citecolor=red
]{hyperref}

\DeclareUnicodeCharacter{0308}{\"{}}

\begin{document}

\title{MXene with Janus Structure at Transition metal site –A route to Emergent Properties} 
\author{Rajdeep Biswas}
 \email{biswasprabal49@gmail.com}
 \affiliation{Department of Condensed Matter and Materials Physics,
S. N. Bose National Centre for Basic Sciences, Kolkata 700106, India}
\author{Tanusri Saha Dasgupta}
 \email{t.sahadasgupta@gmail.com}
 \affiliation{Department of Condensed Matter and Materials Physics,
S. N. Bose National Centre for Basic Sciences, Kolkata 700106, India}
\date{\today}

\begin{abstract}
Motivated by the discovery of bimetallic MXene compounds with Janus metal sites, we investigate Janus MXenes TiM$^{''}$CO$_2$, where M$^{''}$ = Mo, W. Our computational analysis reveals that broken inversion symmetry in the Janus structure, coupled with strong spin-orbit coupling at M$^{''}$, generates diverse, remarkable functionalities. These include pronounced Rashba effect, non-trivial Z$_2$ topology, Berry curvature dipole-driven non-linear anomalous Hall effect, and strain-control of Berry curvature dipole. Notably, 4d transition metal-based TiMoCO$_2$ and 5d transition metal-based TiWCO$_2$, with M$^{''}$ from the same column of the periodic table, display markedly different behaviors. While TiMoCO$_2$ acts as a Z$_2$ topological insulator, TiWCO$_2$ functions as a trivial semi-metal. Both compounds, however, display compelling quantum properties. TiWCO$_2$ shows a robust Rashba effect with a large Rashba coefficient of
1.35 eV\AA\ and a colossal non-linear Anomalous Hall conductivity of 120 × 10$^{-4}$ $G_0$. TiMoCO$_2$, a Z$_2$ narrow-gap semiconductor with weaker Rashba effect and moderate non-linear Anomalous Hall conductivity, exhibits a strain-driven transition from semiconductor to semi-metal. This transition modulates both the sign and magnitude of the Berry curvature dipole, yielding a sizable nonlinear anomalous Hall conductivity of 17 × 10$^{-4}$ $G_0$ under 2$\%$ tensile strain. Our findings underscore the potential of MXene as a platform for investigating and tailoring quantum functionalities.
\end{abstract}
\maketitle

\section{Introduction}

Novel properties exhibited by two-dimensional (2D) materials have  become one of the most exciting areas of current interest in science and technology\cite{berman2024introduction}. Since the discovery of single-layer carbon compound of graphene, the list of 2D materials is ever-growing, with the inclusion of compounds like boron nitride (BN), transition-metal dichalcogenides (MoS$_2$, WS$_2$, WTe$_2$),  phosphorene, etc.\cite{manzeli20172d,vatanpour2021comprehensive,carvalho2016phosphorene}. A recent addition to this excitement is 2D transition-metal carbide and/or nitride layers, so-called MXenes\cite{chakraborty20191}. MXenes are chemically exfoliated from three-dimensional (3D) layered MAX compounds of general formal M$_{n+1}$AX$_n$ where $n$ = 1, 2, or 3, “M” is an early transition metal like Sc, Ti, Zr, Hf, V, Nb, Ta, Cr, or Mo, “A” is an element from groups III-VI in the periodic table (Al, Ga, In, Tl, Si, Ge, Sn, Pb, P, As, Bi, S, or Te), and “X” is carbon and/or nitrogen. Depending on the choice of transition metals, thickness, and surface functionality, 2D MXene compounds of general formula M$_{n+1}$X$_n$T$_2$ (T being the surface terminating unit) may exhibit unique properties that are not shared by the parent MAX phases\cite{khazaei2019recent}. 

There exists experimental reports\cite{naguib2012two,yang2016two} supporting that alloying at transition metal sites (M$^{'}$ and M$^{''}$) as well as “A” sites (A$^{'}$ and A$^{''}$) or “X”  sites (C and N) is possible,  resulting in the formation of alloy MAX phases and consequently the synthesis of various alloy MXenes. Interestingly, by selection of elements and stoichiometry, they can be formed even in an ordered state \cite{anasori2015experimental,anasori2015two,meshkian2017theoretical}. Focusing on transition metal sites with 50:50 composition, it may exist as an out-of-plane ordered double transition metal form, known as o-MXene or as disordered with a mix of M$^{'}$ and M$^{''}$ in each layer, known as i-MXene. Both forms, namely, i-Mxene and o-MXene, have been synthesised \cite{lu2017theoretical,meshkian2018w,chen2018theoretical,dahlqvist2018origin,halim2018synthesis}. For $n$ = 1 MXene, with two transition metal layers, out-of-plane ordering leads to the Janus structure of M$^{'}$M$^{''}$X with broken inversion symmetry, where one of the layers of the MXene consists of transition metal M$^{'}$ only and the other layer consists of transition metal M$^{''}$ only.  This is in parallel to Janus structures formed out of transition metal dichalcogenides (TMDC) MXY, e.g. WSSe or MoSSe \cite{tang20222d}, with the role of chalcogens played by transition metals and W/Mo played by C/N. Janus transition metal dichalcogenides have been studied\cite{tang20222d,cocskun2025experimental,kaneda2024nanoscrolls} for their attractive properties like Rashba effect, Berry curvature dipole, etc. On the other hand, similar studies for the Janus MXene structure are limited \cite{PhysRevB.107.075403}. MXenes, on the other hand, offer several advantages over TMDs due to their possible metallic conductivity, tunable surface functionality through the terminating group, and better cycling stability \cite{shahzad20202d,hu2019ordered,PhysRevMaterials.4.124007}.

\begin{figure}
    \centering
    \includegraphics[width=1.0\linewidth]{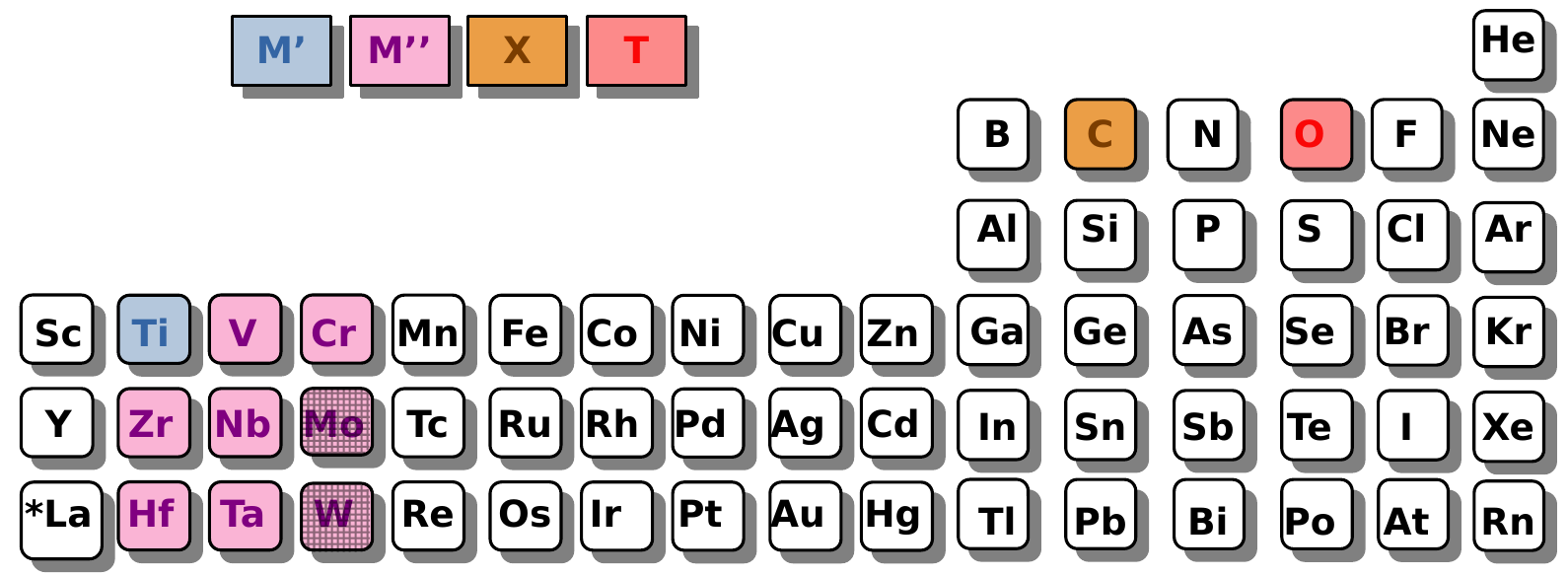}
    \caption{\protect\justifying Periodic Table showing the various different elements of bimetallic M$^{'}$M$^{''}$XT$_2$ MXenes, considered in present study. M$^{'}$, X and T are fixed as Ti, C and O in 
    all compounds, while M$^{''}$ is varied over different early transition metals from 3d (V, Cr) series, 4d (Zr, Nb, Mo) series and 5d (Hf, Ta, W) series. The M$^{''}$ elements forming 
    TiM$^{''}$CO$_2$ Janus o-MXene structures are shown with hatches.}
    \label{fig:figure1}
\end{figure}

In this study, our aim is to computationally explore the emergent properties that a monolayer Janus MXene may offer by combining two essential ingredients: a) broken inversion symmetry and b) strong spin-orbit coupling (SOC) at the transition metal site. Starting with a prominent member of the MXene family, Ti$_2$CO$_2$, consisting of a carbon layer sandwiched between two titanium layers, with oxygen atoms terminating the surface, we consider alloying with early transition metals of 3d series (V,Cr), 4d series (Zr, Nb, Mo) and 5d series (Hf, Ta, W) (cf Fig.~\ref{fig:figure1}). Among the various different (Ti, M$^{''}$) combinations studied, TiMoCO$_2$ and TiWCO$_2$ are found to form the out-of-plane ordered  o-Mxene Janus structure with broken inversion symmetry and strong SOC at 4d Mo or 5d W site.  Interestingly, TiMoCO$_2$ and TiWCO$_2$ exhibit contrasting properties. While surface passivation by O makes TiMoCO$_2$ semiconducting as in its parent structure of Ti$_2$CO$_2$, TiWCO$_2$ turns out to be metallic. As opposed to trivial metallic properties of TiWCO$_2$, TiMoCO$_2$ is found to be a Z$_2$ topological insulator with topologically protected surface states. The effect of 
strong spin-orbit coupling in the broken inversion symmetry set-up manifests in multiple quantum functionalities in two compounds. TiWCO$_2$ turns out to be a strong Rashba compound, hosting 
an unprecedentedly large value of Berry curvature dipole and consequently colossal non-linear anomalous Hall effect (NLAHE), making
the situation appealing with a combination of several emergent properties: metallicity, large Berry curvature dipole and strong Rashba effect. TiMoCO$_2$, which is a non-trivial narrow-gap semiconductor, shows remarkable modulation of its electronic properties under modest tensile strain. Application of strain on 
TiMoCO$_2$ is found to drive a phase transition from semiconductor to semi-metallic electronic state, maintaining the non-trivial topology. The change in band structure properties under strain allows
for strain-tuning of non-linear Hall current through control of the magnitude and sign of Berry curvature dipole, promising high-efficiency terahertz detectors, switches, and high-sensitivity devices \cite{sodemann2015quantum,ma2019observation,liang2021janus,kang2019nonlinear,xiao2012coupled}.
This, in turn, opens up a wide canvas of intriguing and technically promising properties in Janus MXene, which is  expected to stimulate further investigation in this relatively unexplored territory of MXene.

\section{Computational Methods}

First principle density functional theory (DFT) calculations with choice of projector augmented-wave pseudopotentials and Perdew-Burke-Ernzerhof (PBE) generalized gradient approximation (GGA) as exchange correlation functional\cite{perdew1996generalized} are carried out as implemented in the plane-wave based Vienna ab initio simulation package (VASP)\cite{hafner2008ab}. The convergence in  self-consistent field (SCF) calculations
is achieved with plane-wave cutoff energy of 600 eV. A large vacuum space of $\sim$ 15\AA\ is used to minimize the artificial interaction between periodic images of the 2D MXene within 3D periodic set-up of the calculations. Structural optimizations are carried out with respect to internal atomic coordinates and unit cell volume with convergence threshold of 10$^{-5}$ eV for total energy and 10$^{-3}$ eV/\AA\ for maximum force/atom employing 10$\times$10$\times$1 Monkhorst-Pack mesh\cite{monkhorst1976special}. Converged k-mesh of 15$\times$15$\times$1 is used for SCF calculation with tight energy convergence threshold of 10$^{-6}$ eV. 
The electronic structures and derived topological and transport properties discussed in this work were calculated within the nonmagnetic PBE+$\mathrm{SOC}$ framework. The relativistic spin orbit coupling (SOC), which is important for 4d and 5d heavy transition metals, is included in all the calculations. 

To study topological properties such as Berry curvature of the constructed Janus structures, maximally localized Wannier functions (MLWFs) for each structures are computed to derive a tight-binding model from ab initio calculations taking different orbital basis using WANNIER90 code\cite{w90}. To check the non-triviality we calculate the Z$_2$ topological invariant following the procedure by Soluyanov et al,\cite{soluyanov2011computing} using the notion of Wannier charge centers (WCCs). 

Although the non-linear effect is related to the net Berry curvature due to a nonequilibrium Fermi distribution, this can be described by Berry curvature dipole defined in the equilibrium state in the semiclassical approximation. In particular, the nonlinear conductivity $\chi_{abb}$ can be expressed as follows,

\begin{equation}
\chi_{abb} = -\epsilon_{adb} \dfrac{e^3\tau}{2\hbar^{2}(1+i\omega\tau}) D_{bd}
\end{equation}

\begin{equation}
D_{bd} = \int_{k} f^{0}_{n}(\vec{k}) \dfrac{\delta\Omega_{d}^{n}}{\delta k_{b}}
\end{equation}

where $D_{bd}$ is the BCD, $f^{0}_{n}(\vec{k})$ is the equilibrium Fermi-Dirac distribution, $\tau$ is the relaxation time, and $\epsilon_{adb}$ is the third rank Levi-Civita symbol ($a, b = x, y$ and $d = z$) in 2D. We compute $\Omega_{d}$ and then calculate the $D_{bd}$ by integrating $\dfrac{\delta\Omega_{d}^{n}}{\delta k_{b}}$ in a very dense $k$-grid of 1000 $\times$ 1000 $\times$ 1 to obtain converged values of the Berry curvature dipole. We employ WANNIER-BERRI code\cite{tsirkin2021high} for this purpose. Note, the Berry curvature dipole is dimensionless in three dimensions, whereas it is in unit of length in 2D. 
Details of structural stability calculations in terms of phonons, ab initio molecular dynamics simulations are given in the Supplemental Material (SM)\cite{SM}.

\section{Results and Discussions}

\subsection{Energetics of o-MXene vs i-MXene}

\begin{figure*}
    \centering
    \includegraphics[width=0.7\linewidth]{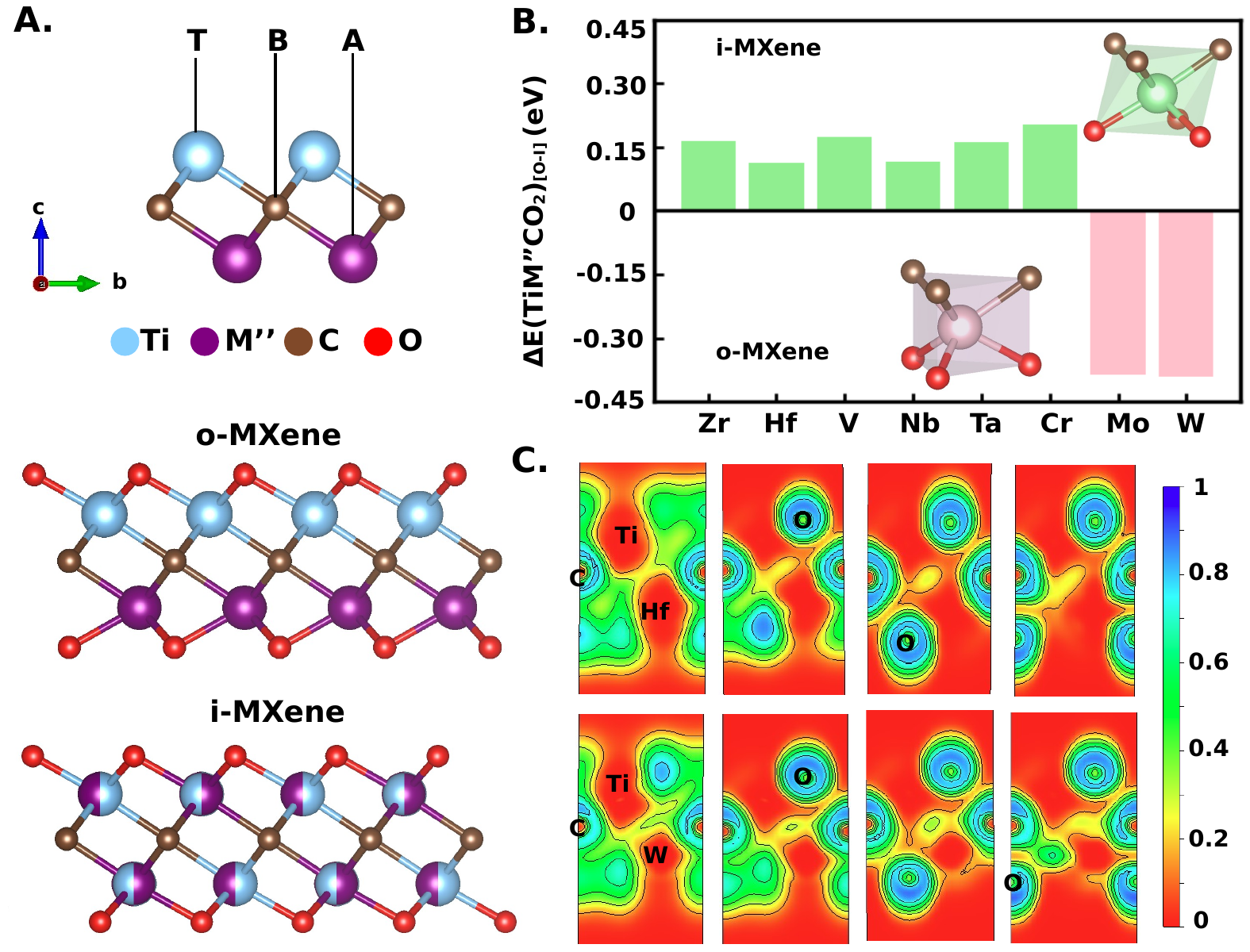}
    \caption{\protect\justifying{\bf A.} Top: Available absorption sites for passivating O atom, T, B and A (see text for details), and Bottom: TiM$^{''}$CO$_2$ in Janus o-MXene  and 
    mixed i-MXene structures. {\bf B.} Energy difference ($\Delta$E(TiM$^{''}$CO$_2$)$_{o-i}$) between o-MXene and i-MXene configurations for different (Ti, M$^{''}$ = Zr, Hf, V, Nb, Ta, Cr, Mo, W) combinations. Inset shows the octahedral A-passivation (top) and trigonal prismatic B-passivation (bottom) coordination formed at C/O atoms around the metal sites (Ti or M$^{''}$), upon O absorption at A and B sites, respectively. {\bf C.} 2D slices projected along [110] direction of ELF for (Ti, Hf) (upper) and (Ti, W) (lower) MXenes. From left to right the panels
    shows the ELF plots for TiM$^{''}$C, TiM$^{''}$CO,  AA-TiM$^{''}$CO$_2$, and AB-TiM$^{''}$CO$_2$.}
    \label{fig:figure2}
\end{figure*}

In order to consider MXene alloy structures constructed out of M$^{'}$ and M$^{''}$ transition metal elements, we consider two different possible structures, o-MXene and i-MXene. An unpassivated  layered structure of M$_{2}$C MXene consists of three atomic layers in a trigonal lattice in which layer of C atoms is sandwiched between two metal (M) layers. For o-MXene, the top (bottom) metal layer is occupied solely by M$^{'}$ (M$^{''}$) atoms, resulting in an inversion symmetry broken structure. For i-MXene, on the other hand, both top and bottom layers are of mixed composition M$^{'}_{0.5}$M$^{''}_{0.5}$, resulting in inversion-symmetric structure (see Fig. ~\ref{fig:figure2}A).  For i-MXene, with mixed composition of metal layers, several different configurations are possible. Thus a configuration  averaging was carried out considering all possible distinct configurations within a supercell of 2 $\times$ 2 $\times$ 1. For details see SM \cite{SM}. The energetics also crucially depends on the site selectivity of  O passivation. The MXene surface has three distinct adsorption sites, as shown in top panel of  Fig.~\ref{fig:figure2}A,  one at the top of the metal atom in the top layer (T), second one at the top of the C atom in the middle layer (B) and the third one  at the top of the metal atom in the bottom layer (A). Out of these, the T-passivation is energetically much higher than B and A due to steric hindrance. Whether the B or A is preferred depends on the type of metal and interplay of C-O covalency and M-O covalency. Upon absorption of O at B site, the metal atom forms a trigonal prismatic coordination with 3 C atoms and 3 O atoms with {\it hcp} stacking, while for absorption at A site, an octahedral coordination of C/O atoms around the metal atom with {\it fcc} stacking is formed (cf insets in Fig. ~\ref{fig:figure2}B). Interestingly, as shown in Fig. ~\ref{fig:figure2}B, while for group 4B and 5B transition metals (Ti, V, Zr, Nb, Hf, Ta) A site is preferred for O passivation, for group 6B transition metals (Cr, Mo, W) B site is preferred. This is explainable by the plot of  electron localization function (ELF) analysis \cite{savin1992electron,savin1997elf}, shown in Fig. ~\ref{fig:figure2}C for the representative cases of (Ti,Hf) and (Ti,W).  The value of ELF between two atoms, which measures the localization of pairs of electrons,  can be in the range of 0 to 1, where 1, 0.5, and 0 represent covalent, metallic, and non-bonding character, respectively. The top panels of Fig. ~\ref{fig:figure2} show the ELF plots for TiHfCO$_2$ for unpassivated TiHfC, TiHfCO with O passivation at A site of Ti layer, TiHfCO$_2$ with O passivation at AA configuration of top Ti layer and bottom Hf layer, and TiHfCO$_2$ with O passivation at AB configuration of top Ti layer and bottom Hf layer. The bottom panels show the same for TiW MXene.  AA (AB) configuration is preferred for TiHfCO$_2$ (TiWCO$_2$). Interestingly, after the absorption of the first O at Ti side, a highly localized ELF appears over A and B site in TiHfCO and TiWCO, respectively, establishing the electronic origin of preferred formation of AA and AB site passivation of (Ti,Hf) and (Ti,W) in O-MXene configuration. For i-MXene, with mixed Ti and M$^{''}$ layers, for all studied choices of M$^{''}$ AA passivation is favoured, which establishes the deciding effect of Ti. Fig ~\ref{fig:figure2} B. shows the energy difference between TiM$^{''}$CO$_2$ in o-MXene and i-MXene configuration for M$^{''}$ = V, Cr, Zr, Nb, Mo, Hf, Ta and W, in their minimum energy AA (see upper inset) or AB configuration (see lower inset) of O passivation. As found, while apart from TiMoCO$_2$ and TiWCO$_2$, all other Ti and M$^{''}$ combinations lead to preference  of i-MXene over o-MXene. The preference to o-MXene of TiMoCO$_2$ and TiWCO$_2$ is guided by the large electronegativity difference, as presented in SM\cite{SM}.

\begin{figure*}[ht!]
    \centering
    \includegraphics[width=0.75\linewidth]{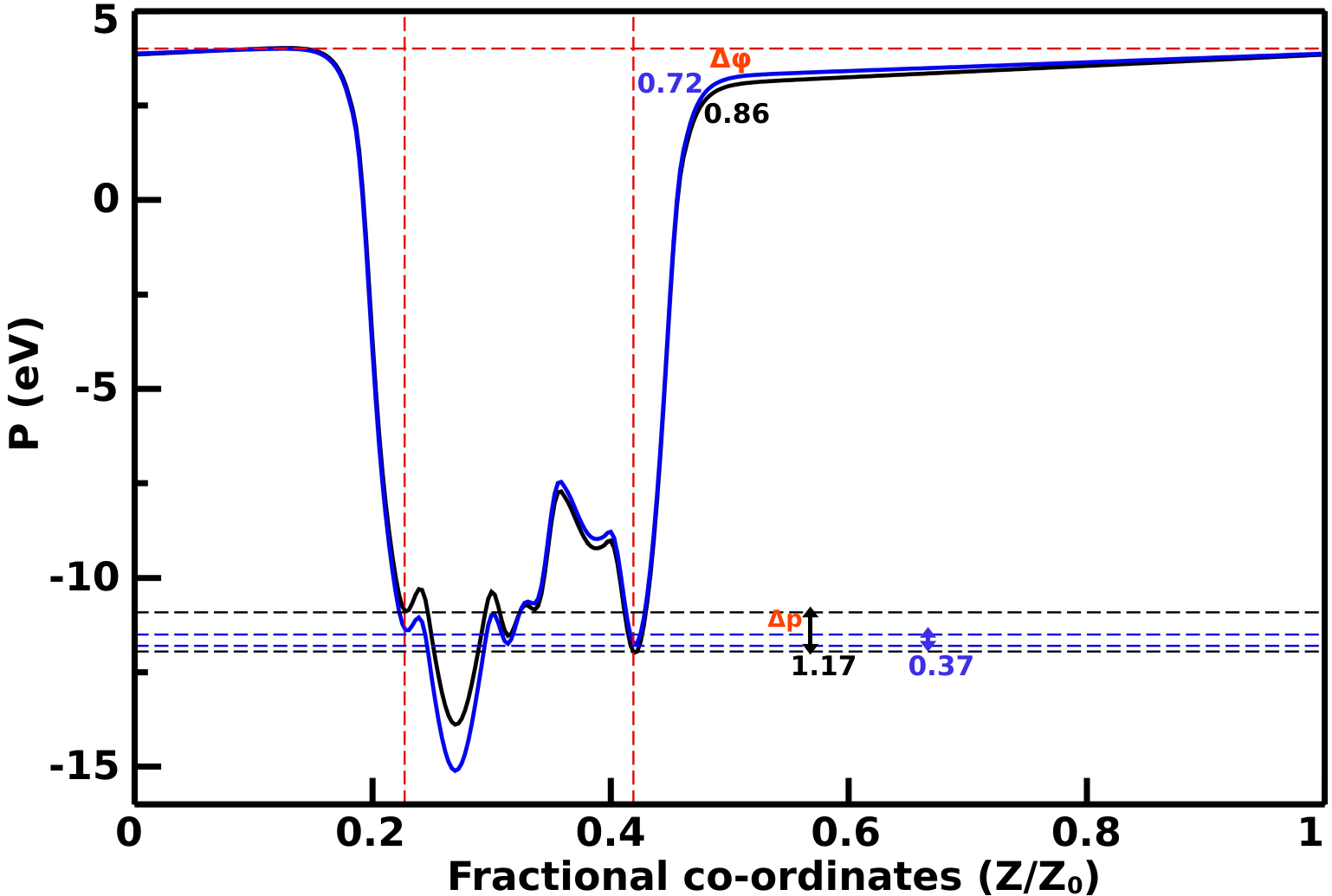}
    \caption{\protect\justifying Planar average of the electrostatic potential of Janus TiMoCO$_{2}$ (black)
    and TiWCO$_2$ (blue)  MXenes along the out-of-plane direction. Marked are $\Delta p$ and $\Delta \phi$.}
    \label{fig:figure3}
\end{figure*}

\begin{figure*}[ht!]
    \centering
    \includegraphics[width=1\linewidth]{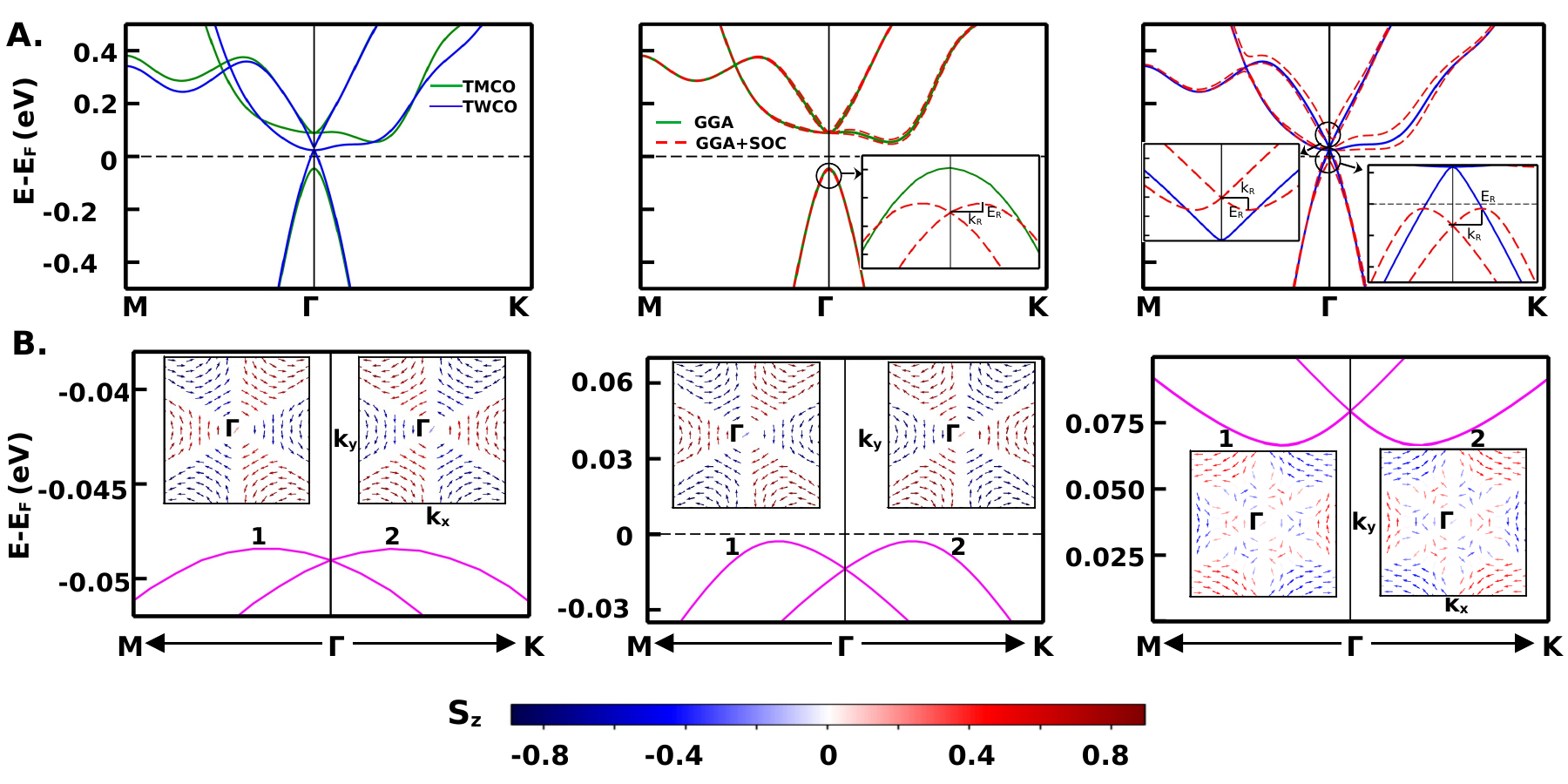}
    \caption{\protect\justifying{\bf A.} GGA band structure of TiMoCO$_2$ and TiWCO$_2$ (left), comparison of GGA and GGA+SOC band structure of TiMoCO$_2$ (middle), comparison of GGA and GGA+SOC band structure of TiWCO$_2$ (right) plotted along the high-symmetry points, M(0.5, 0, 0), $\Gamma$(0, 0,0), and K(0.333, 0.333, 0) of the trigonal BZ.  Insets show the Rashba crossing close to Fermi level. Marked are E$_R$ and k$_R$ (see text for details). {\bf B.} The Rashba spin textures corresponding to Rashba split bands (marked as "1", shown in left of the panel and as "2", shown in right of the panel) for crossing below E$_F$ in TMCO (left), below E$_F$ (middle) and above E$_F$ (right) in TWCO.}
    \label{fig:figure4}
\end{figure*}

Since we are interested in Janus structures, in the following we focus only on o-MXene structured TiMoCO$_2$ (TMCO) and TiWCO$_2$ (TWCO). The thermal and dynamical stability
TMCO and TWCO in o-MXene structure are ensured through ab-initio molecular dynamics calculations at 300 K temperature and phonon dispersions. See SM\cite{SM}.

\subsection{Polarization Properties}

Inversion symmetry breaking in o-MXene-structured TiMoCO$_2$ and TiWCO$_2$ gives rise to an internal
electric-field generating dipole moment. Fig. 3 shows the planar average of the electrostatic
potential of TiMoCO$_2$ and TiWCO$_2$, plotted along the out-of-plane (z) direction.  As in seen, even if the outermost layer
is symmetric, both passivated by O atoms, due to the asymmetric positioning of Ti and Mo/W between the top and bottom metal layers,
sandwitching the C layer, a finite  potential difference ($\Delta p$) of 1.17 eV for TMCO and of 0.37 eV for TWCO appears between the two edges,
resulting in a work function difference ($\Delta \phi$) of 0.86 eV for TMCO and of 0.72 eV for TWCO. The corresponding dipole moment of TMCO and TWCO is found to be 0.0469 e\AA \enskip and 0.036 e\AA, respectively. The weakening of $\Delta p$ resulting in a weaker dipole moment in TWCO compared to TMCO
is rationalised by the stronger covalency of the W-C bond compared to the Mo-C bond. The electronegativity difference of W and C is $\sim$ 0.19
compared to $\sim$ 0.39 for Mo and C. W 5d orbitals are more extended compared to Mo 4d, and thus having 
stronger hybridization with C 2p, which lead to broader bands and stronger electronic de-localization. 

This difference in covalency is reflected in the GGA band structure of TMCO and TWCO, which shows the semiconducting nature of TMCO with an indirect gap of $\sim$ 0.1eV
and a metallic character of TWCO (cf left most panel in Fig. ~\ref{fig:figure4} A). Inversion symmetry broken structure upon inclusion of spin-orbit coupling (SOC) shows Rashba-like
splitting at $\Gamma$ point close to Fermi level (E$_F$) for the valence band in narrow-gap semiconductor TMCO (cf middle panel in Fig.  ~\ref{fig:figure4} A), and for below and above E$_F$ for metallic TWCO 
(cf right-most panel in Fig.  ~\ref{fig:figure4} A). The spin degeneracy is lifted except for the crossing point
of two branches (Dirac point). The Rashba parameter, $\alpha_R$ is defined as  $\frac{2E_R}{k_R}$, where the Rashba energy $E_R$ is 
defined as the energy difference between the extrema of the split band and the Dirac point, while $k_R$ is the minimum of
the lower branch in momentum space measured from the crossing point. Following this definition, TMCO shows $\alpha_R$ of 0.23 eV\AA \enskip at an energy
$\sim$ 50 meV  below E$_F$, while TWCO shows $\alpha_R$ = 1.06 eV\AA \enskip  at an energy
of $\sim$ 15 meV  below E$_F$ and another of $\alpha_R$ = 1.34 eV\AA \enskip at $\sim$ 80 meV  above E$_F$.  The stronger value of $\alpha_R$ reflected in the stronger spin splitting of bands in TWCO compared to TMCO is due to the stronger SOC of 5d W compared to 4d Mo. This characterizes TWCO as a strong Rashba compound following 
the definition by Acosta et al.\cite{acosta}.  Such metallic Rashba system can be useful in inducing Rashba
SOC in non-metallic hexagonal monolayers like graphene or transition-metal dichalogenide through proximity effect, which has become  an important research direction lately for spintronic device fabrication\cite{xiao2012coupled,zhu2011giant,shukla2025co}.

The Rashba spin texture around the crossing point for TMCO valence band, and for TWCO below and above E$_F$, is shown in Fig. ~\ref{fig:figure4} B, which encodes
the characteristic tangential arrangement of electron spins in momentum space, with spins rotating in opposite directions—one clockwise and the other counterclockwise, corresponding to
Rashba split bands.

\subsection{Topological Properties}

\begin{figure*}[ht!]
    \centering
    \includegraphics[width=0.7\linewidth]{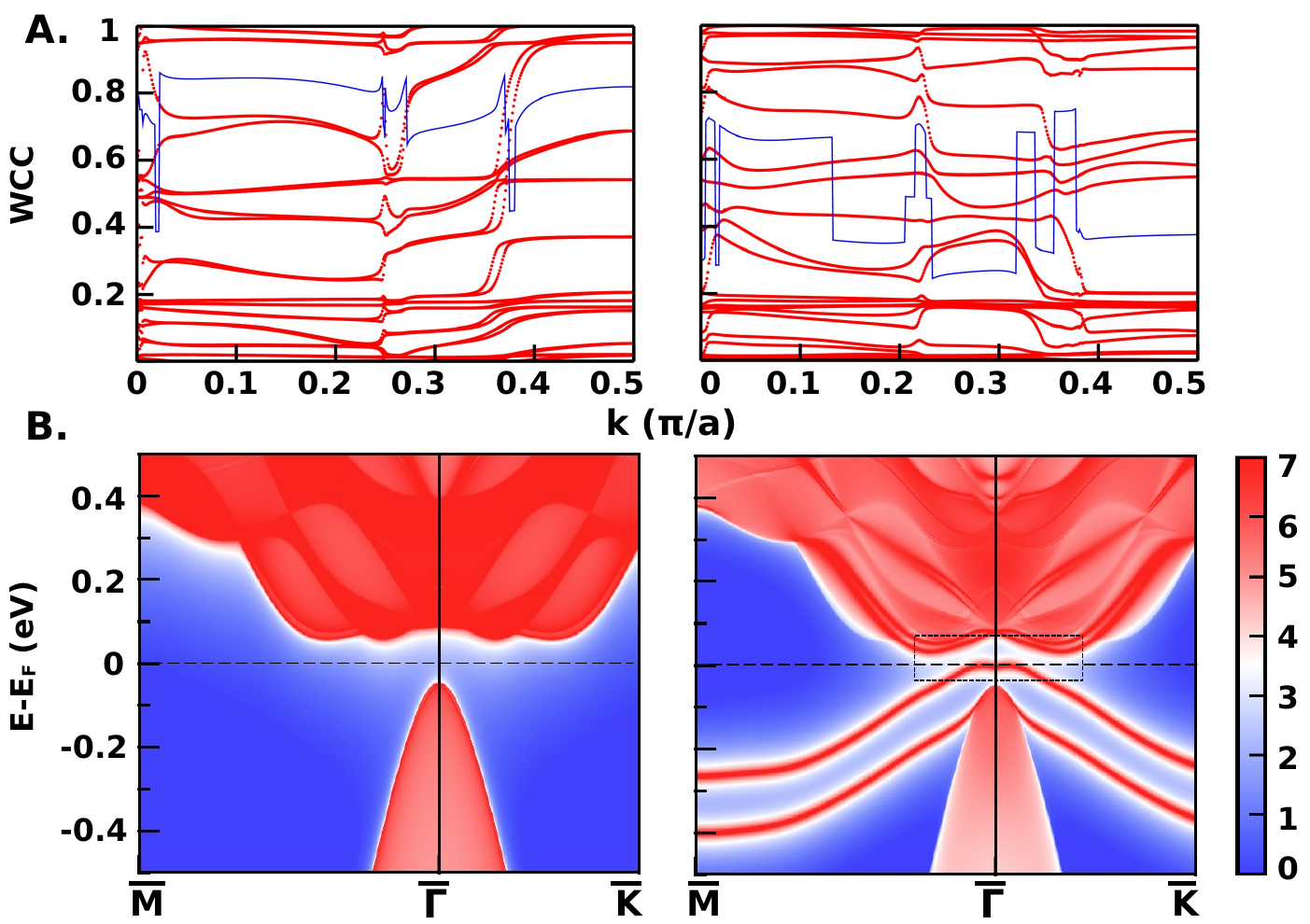}
    \caption{\protect\justifying{\bf A.} The WCCs tracked along the $k_z$ = 0 plane for
TiMoCO$_2$ (left) and TiWCO$_2$ (right). The blue line indicate the maximum interspace function $\zeta$(k), which intersects the WCCs (red
dotted) an odd number of times for TiMoCO$_2$ and even number of times for TiWCO$_2$, 
indicating that topological non-triviality and triviality of the compounds, respectively. {\bf B.} Spectral function plot for the bulk (left) and combined
edge and bulk (right) (in [010] direction) of Z$_2$ topological semiconductor TiMoCO$_2$. The conducting edge
spectra has been highlighted by the box. The zero of the
energy scale is set at the Fermi level.}
    \label{fig:figure5}
\end{figure*}

Presence of strong SOC coupled with broken inversion symmetry may give rise to interesting topological properties in the two o-MXene
structures under study. To probe that, we first calculated the Z$_2$ topological invariant. The method proposed by Soluyanov et al.\cite{soluyanov2011computing} based on the notion of Wannier charge centers (WCCs), which is the expectation value of the position operator in the state corresponding to the Wannier functions of the unit cell, is used. The evolution of the WCCs among the time reversal invariant plane $k_z$ = 0 of the nonmagnetic ground state of TMCO and TWCO is shown in Fig ~\ref{fig:figure5}A.  A Kramers pair is topologically non-trivial if the corresponding WCC curves at $k$ = 0
and $k$ = $\pi$ belong to a different branch. This topological state can be verified by tracking the discontinuities in the maximum interspace between WCCs, denoted as 
($\zeta$(k)) rather than curves themselves. Z$_2$ is given by the number of discontinuities of $\zeta$(k) in [0, $\pi$]
modulo 2. As found in Fig.  ~\ref{fig:figure5} A, the $\zeta$(k) function (blue line) drawn parallel
to the k axis intersects the WCCs (red line) an odd number of times for TMCO and an even number of times for TWCO, 
indicating the contrasting topological properties of TMCO and TWCO, the former being topologically non-trivial and the latter being trivial.

To further substantiate the topological nature of TMCO, which should result in topologically protected edge states, we computed the 
spectral function using the surface Green’s function formalism, shown in Fig.  ~\ref{fig:figure5} B, as implemented in WannierTools \cite{wu2018wanniertools}. The color intensity in the plots reflects the degree of surface localization of electronic states. The shaded regions indicate the bulk band projections of 
the slab calculation, and the sharp, intense lines within these regions are the true edge states. A well-defined conducting edge state is 
clearly observed along the [010] direction, appearing at the $\bar{\Gamma}$ point. The edge states, protected by time-reversal symmetry, are robust against nonmagnetic impurities and disorder.
TMCO is thus characterized as Z$_2$ topological insulator.

\begin{figure*}[ht!]
    \centering
    \includegraphics[width=0.8\linewidth]{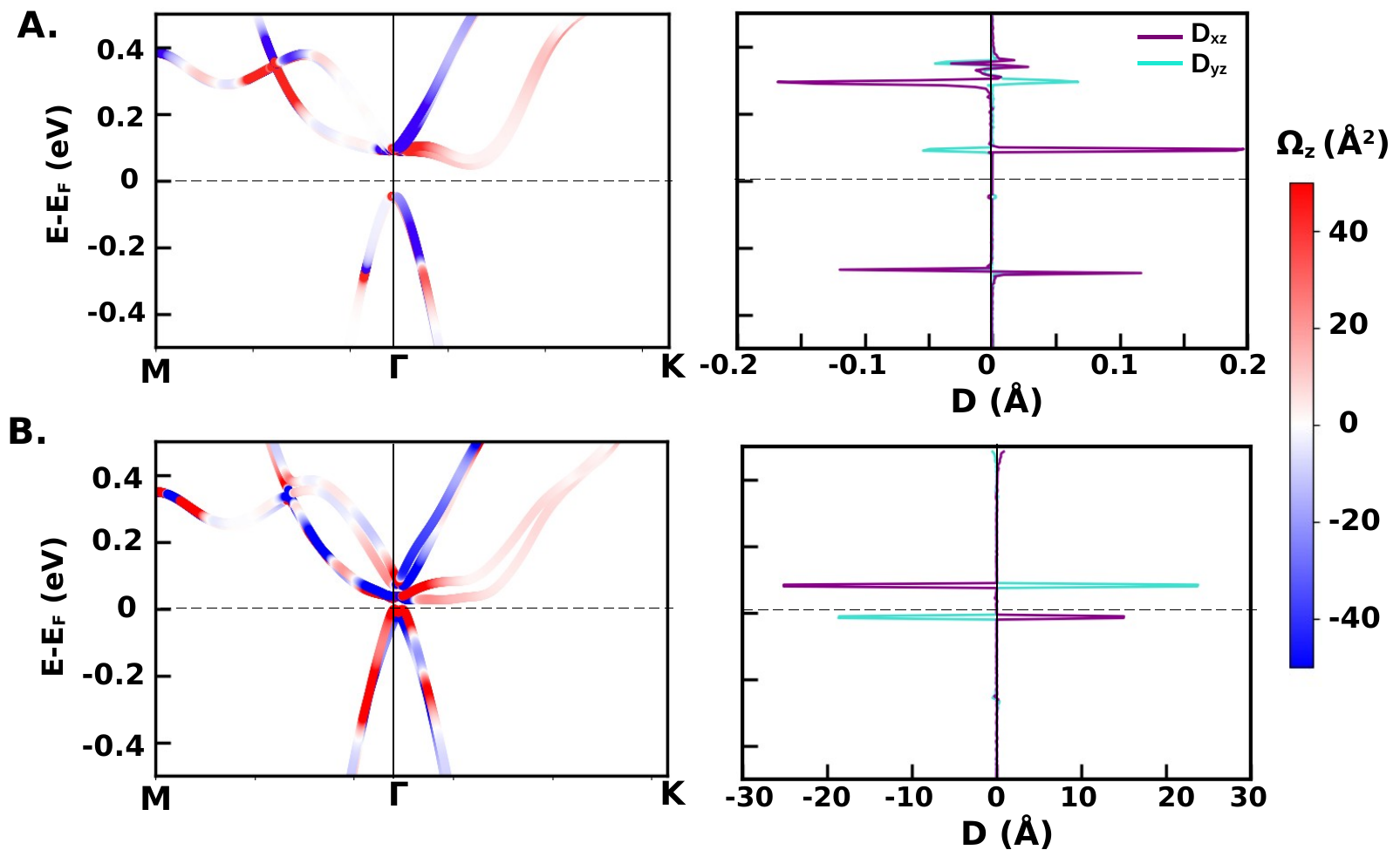}
    \caption{\protect\justifying{\bf A.} Distribution of Berry curvature plotted along high-symmetry direction for TiMoCO$_2$ (left). Corresponding Berry curvature dipoles, $D_{xz}/D_{yz}$ plotted as a function of 
    energy measured with respect to Fermi level (right). {\bf B.} The same, but plotted for TiWCO$_2$.}
    \label{fig:figure6}
\end{figure*}

Additionally, though net Berry curvature is zero due to presence of time reversal symmetry, absence of inversion symmetry in nonmagnetic Janus TMCO and TWCO can give rise to Berry curvature dipole leading to anomalous velocity of Bloch
electrons and nonlinear Hall response\cite{du2021quantum}. In 2D crystals, Berry curvature  behaves as a pseudoscalar and has
a nonzero out-of-plane component ($\Omega_z$), and its first-order moment, the Berry curvature dipole, behaves as a pseudovector $\vec{D}$ 
in the 2D plane with components $D_{xz}$ and $D_{yz}$. Fig.  ~\ref{fig:figure6} A shows the distribution of Berry curvature along the high-symmetry direction of BZ
for TMCO and TWCO. As is seen, Berry curvature is  strong localized in momentum space, especially near band crossings and anti-crossings,
both for TMCO and TWCO. Further, the Berry curvature hot-spot gets shifted or tilted from the high-symmetry $\Gamma$ point, reflecting the broken
inversion symmetry. These give rise to sharply peaked Berry curvature dipole (BCD) in the energy range of $\pm$0.4 eV, both for TMCO with 
non-trivial band topology, and TWCO with trivial band topology. While for TMCO, the BCD ($D_{xz}/D_{yz}$) values turn out to be 0.1-0.2 \AA\,
comparable to that reported for MoS$_2$ and WTe$_2$\cite{fs39-vgq6,joseph2021tunable}, BCD takes extremely large values of 24-25 \AA\ for TWCO, which are about two orders
of magnitude larger than that for TMCO. Taking electric field along $x$/$y$ direction (E$_x$/ E$_y$), 
the non-linear Hall conductivity can be computed using the formula: $\sigma_{yx}=\chi_{yxx}E_x=\dfrac{e^3\tau E_x}{\hbar^2}D_{xz} = G_0\dfrac{\tau D_{xz} e E_x  \pi}{\hbar}$;
$\sigma_{xy}=\chi_{yxx}E_y =\dfrac{e^3\tau E_y }{\hbar^2}D_{yz} = G_0\dfrac{\tau D_{yz} e E_y  \pi}{\hbar}$
where $G_0 = \dfrac{2e^2}{h}$ is the conductance quantum.

Considering the applied field, $E_{x}/E_{y}$=10$^3$ V/m and $\tau$=10$^{-12}$sec, the non-linear conductivity 
corresponding to $D_{xz}$ ($D_{yz}$) peak for TMCO and TWCO are obtained as, 0.95 (0.34) $\times$ 10$^{-4}$ $G_0$ and 120 (113) $\times$ 10$^{-4}$ $G_0$. 
This establishes TWCO to be a potential candidate for Quantum frequency doubling, having a nonlinear conductivity about two orders stronger than current WTe$_2$ literature benchmarks \cite{wang2019ferroelectric}.
Metallic TWCO, thus, offers a technologically attractive platform, with strong Rashba effect on one hand and colossal value of BCD enabling high-performance non-linear Hall responses on the other hand.

\begin{figure*}[ht!]
    \centering
    \includegraphics[width=0.8\linewidth]{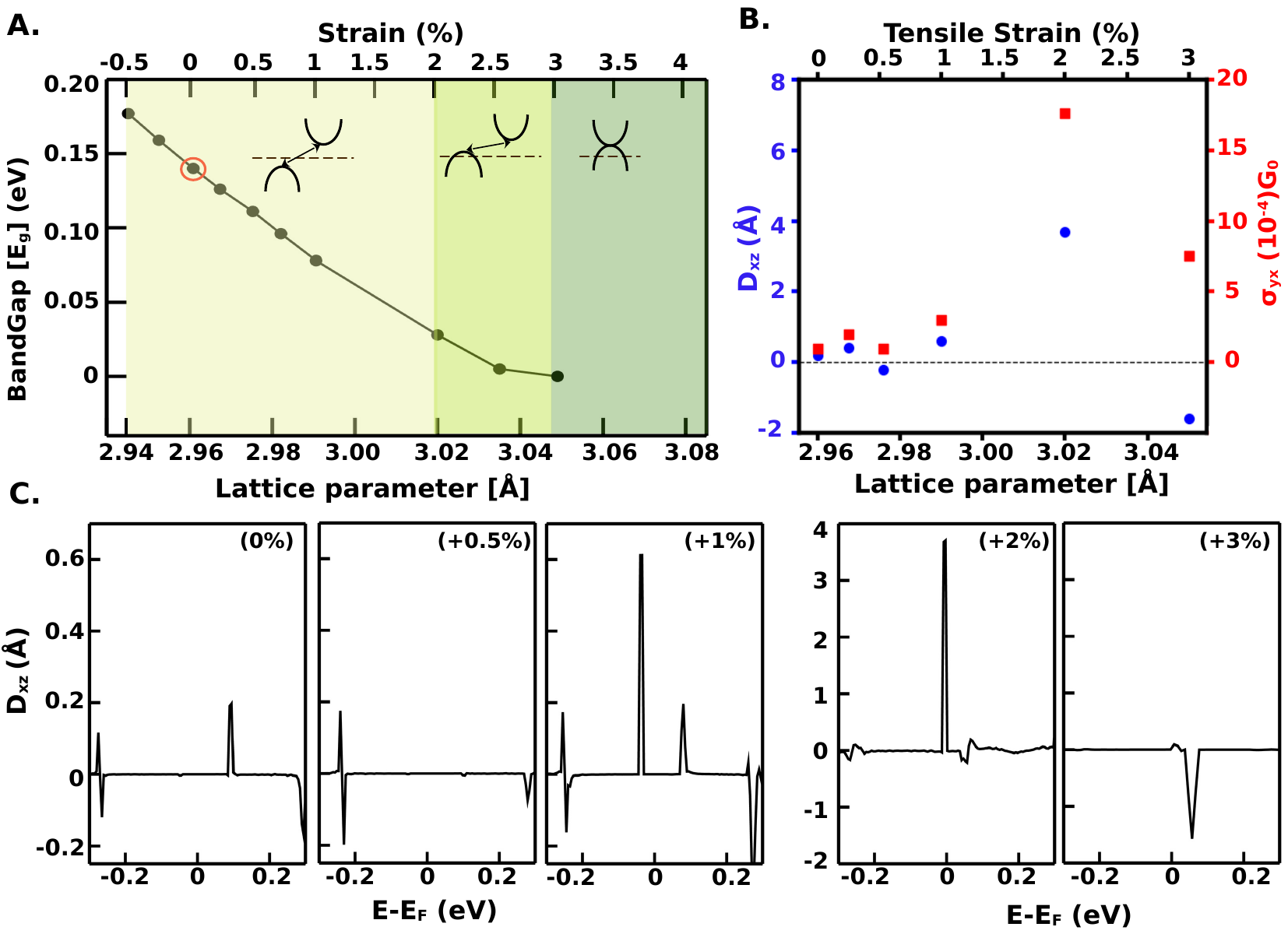}
    \caption{\protect\justifying{\bf A.} The variation of band gap of TMCO under biaxial strain, tensile (compression) strain being denoted as + (-). Around 2-2.5$\%$ tensile strain semiconducting to semi-metallic transition happens in which lower bands cross E$_F$ while beyond 3$\%$ tensile strain, the energy gap between the lower and upper bands vanishes. 
    Insets show the schematics of the representative band structures. {\bf B.} Variation of peak value of Berry curvature dipole ($D_{xz}$) within $\pm$ 200 meV of E$_F$ and NLAH conductivity ($\sigma_{yx}$) as a function of
    tensile biaxial strain. {\bf C.} Energy variation of $D_{xz}$ for different strain values within $\pm$ 200 meV around E$_F$.}
    \label{fig:figure7}
\end{figure*}

\subsection{Straintronics of TiMoCO$_2$}

The integration of straintronics and topology may provide a highly tunable knob to control the electronic phase transition, "strain-switchable" Hall signal through manipulation
of the sign and magnitude of Berry curvature dipole \cite{antonova2022straintronics}. With this goal, we applied tensile strain on TMCO, which in the unstrained condition is a Z$_2$ non-trivial semiconductor with 
a modest value of non-linear Hall conductivity. As is shown in Fig.~\ref{fig:figure7}, application of a strain of 2-2.5 $\%$ drives the semi-conducting system to semi-metallic, where the valence band 
crosses E$_F$, but the valence and conduction bands remain separated by $\sim$ 5 meV indirect gap, making it a p-type metal. Further increase in strain value to $\sim$ 3 $\%$ makes the valence and conduction band touch, as shown in Fig. ~\ref{fig:figure7}A. Throughout all the studied strain values, the compound remains non-trivial Z$_2$ invariant, thus enabling a strain-induced
phase transition from topological insulating phase to Z$_2$ semi-metallic phase. While in the topological insulating phase, the charge carriers are topologically protected edge state electrons; in the Z$_2$ metallic phase, bulk-originated holes as well as topologically protected 
robust edge state electrons contribute to conduction. The edge states are found to be strongly localised and manifests as sharp features in the spectral function, responsible for dissipation-less conduction in contrast to the broader and more diffuse projections of bulk states, as shown in the SM \cite{SM}.

This drives a non-trivial change in band structure details that has a profound influence on the Berry curvature dipole, and thus the non-linear Anomalous Hall 
conductivity. In Fig. ~\ref{fig:figure7}B, the largest value of Berry curvature dipole within $\pm$ 200 meV around E$_F$ and the corresponding non-linear conductivity are plotted as a function of strain. As is seen, strain modulates the sign as well as the magnitude of Berry curvature dipole, especially for a strain of about 2-3 $\%$. About an 
order of magnitude enhancement in $D_{xz}$ and in $\sigma_{yx}$ is achieved upon application of tensile strain of 2-3 $\%$. Increasing strain from 2$\%$ to 3$\%$ flips
the sign of the Berry curvature dipole, keeping its high value, thus allowing control over the direction of the nonlinear Hall current. 

Fig. ~\ref{fig:figure7} C shows the plot of Berry curvature dipole as a function of energy for different strain values. As found, application of tensile strain also modulates the energy  position of the peak of the Berry curvature dipole. In particular, the Berry curvature dipole peaks at 4 meV below E$_F$ for 2$\%$ strain compared to 100 meV above E$_F$ for
the unstrained situation. Thus, strain adds a valuable control knob in tuning the non-linear topological properties of TiMoCO$_2$. Through modulation of the band gap, semi-metallic phases are generated, which significantly amplify Berry curvature fluctuations, leading to large BCD peaks near the Fermi level.

\section{Conclusion}

In conclusion, in search of Janus structured bimetallic  MXene compounds, with broken inversion symmetry and strong SOC, we designed TiM$^{''}$CO$_{2}$ MXene compounds with the choice of M$^{''}$
from 3d, 4d and 5d transition metal series. Out of eight different (Ti,M$^{''}$) combinations, (Ti,Mo) and (Ti,W) are found to stabilise in out-of-plane ordered Janus structure, while other combinations prefer in-plane mixed (Ti,M$^{''}$) configurations. This trend has been rationalised in terms of the electronegativity difference between Ti and M$^{''}$.  
Focusing on TMCO and TWCO, moving from 4d Mo to 5d W at the M$^{''}$ site, a striking contrast of properties is observed. 
The combination of two guiding ingredients, broken inversion and strong SOC, leads to two diverse paths in TMCO and TWCO.  While TMCO is found to 
exhibit Z$_2$ semiconducting properties, TWCO becomes a trivial metal.
TWCO, on the other hand, is characterised as a strong Rashba compound and hosts an extraordinaryly large non-linear Hall response, which is two orders of magnitude larger compared to
that known for Janus transition metal dichalcogenides\cite{joseph2021tunable}. This makes TWCO a viable platform of technological importance, combining metallicity, strong Rashba, and colossal non-linear effects.
TMCO with moderate non-linear Hall response in unstrained condition, shows a strain-induced transition from Z$_2$ semiconductor to Z$_2$ semi-metal, associated with a substantial increase in
non-linear Hall conductivity and flipping sign of Hall current, through strain modulation of Berry curvature dipole. This opens up a playground for quantum functionalities in the landscape of bimetallic 2D MXenes.

\section*{acknowledgements}

T.S.D. acknowledges funding from the J. C. Bose National Fellowship (JCB/2020/000004) for the support during initiation of the project. The authors acknowledge computational support
of National Supercomputing mission.

\section*{conflict of interest}
The authors declare no conflict of interest.

\section*{Supporting Information}

Supporting information includes details on the different configurations of i-MXene and their energetics, the electronegativity difference table, the phonon spectra and ab-initio  MD simulated free energy for TMCO and TWCO, and edge state spectra of strain induced metallic TMCO.

\bibliographystyle{unsrtnat}

\bibliography{Ref}
\end{document}